%% file: main.tex
\documentclass{vgtc}                          % final (conference style)
\graphicspath{{figures/}{pictures/}{images/}{./}} % where to search for the images

\usepackage{times}                     % we use Times as the main font
\usepackage{tabu}                      % only used for the table example
\usepackage{booktabs}                  % only used for the table example
\usepackage{lipsum}                    % used to generate placeholder text
\usepackage{mwe}                       % used to generate placeholder figures

\usepackage{enumitem}  % load only once
\usepackage{wrapfig}
\usepackage[defaultcolor=black]{changes}

\setitemize{noitemsep,topsep=0pt,parsep=0pt,partopsep=0pt}

\usepackage{mathptmx}                  % use matching math font

\usepackage{setspace}
\usepackage[compact]{titlesec}

\onlineid{1820}

\vgtccategory{Research}

\vgtcinsertpkg

\title{Evaluating Replay Techniques for Asynchronous Task Handover \\ in Immersive Analytics}

\author{Zhengtai Gou\textsuperscript{*}\thanks{e-mail: zgou7@gatech.edu. \textsuperscript{*} Authors contributed equally.}\\ %
        \scriptsize Georgia Tech %
\and Junxiao Long\textsuperscript{*}\thanks{e-mail: jlong343@gatech.edu}\\ %
     \scriptsize Georgia Tech %
\and Tao Lu\thanks{e-mail: tlu307@gatech.edu}\\ %
     \scriptsize Georgia Tech %
\and Jian Zhao\thanks{e-mail: jianzhao@uwaterloo.ca}\\ %
     \scriptsize University of Waterloo %
\and Yalong Yang\thanks{e-mail: yalong.yang@gatech.edu}\\ %
     {\scriptsize Georgia Tech}
}

\teaser{
  \centering
  \vspace{-1.5em}
  \includegraphics[width=\linewidth]{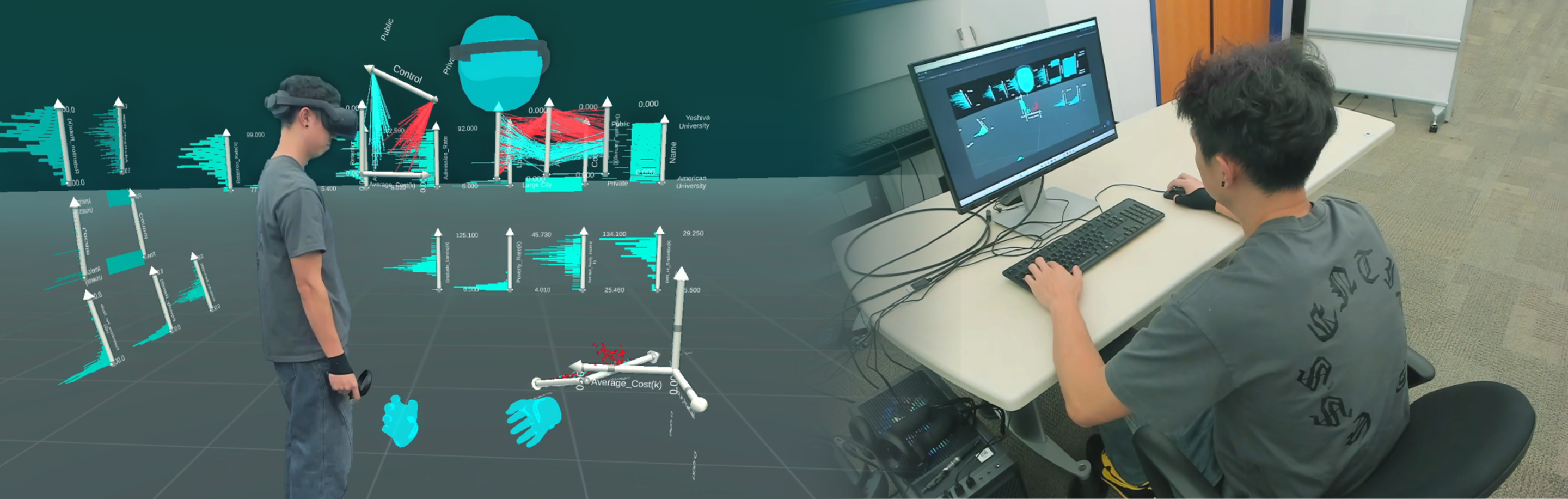}
  \caption{A data analyst (i.e., the real human illustrated in the figure) replays an immersive analytics session, observing the actions of a previous analyst represented by a floating blue avatar. 
  The image depicts the two optimized conditions evaluated in our Phase 2 comparison: the active immersive VR replay and the hybrid desktop PC replay. \vspace{1em}}
  \label{fig:teaser}
}

\input{sections/00-abstract}


\begin{document}



\input{sections/01-intro-v2}

\input{sections/02-related-works-v3}

\input{sections/03-overview}

\input{sections/04-study1}
\input{sections/05-study2}

\input{sections/06-study3}
\input{sections/07-conclusion}

% The authors wish to thank A, B, and C. This work was supported in part by
% a grant from XYZ.}

\acknowledgments{This research was supported in part by NSF award IIS-2441310.}

\thanks{\textit{© 2026 IEEE. Personal use of this material is permitted. Permission from IEEE must be obtained for all other uses, in any current or future media.}}

\bibliographystyle{abbrv}

\bibliography{main}
\end{document}

%% file: sections/00-abstract.tex
\abstract{
     Immersive analytics enables collaborative data analysis in shared virtual spaces. While synchronous collaboration in such environments is well-established, \added{real-world analysis often requires an effective task handover—the transfer of knowledge and analytical context between analysts working asynchronously. Traditional handover methods often rely on static annotations that fail to capture the dynamic problem-solving process and spatial context inherent in immersive workflows. }
    To address this \added{handover challenge}, we explore session replay as a comprehensive approach for analysts to re-experience a predecessor's work, facilitating a deeper understanding of both the visual details and the insight formation process. 
    Two phases of studies were conducted to establish design guidelines for such replay systems by investigating the impact of viewing platform (PC vs. VR), perspective (first-person vs. third-person), and navigation control (active vs. passive). 
    Phase 1 identified the optimal replay configurations within each viewing platform, revealing a platform-dependent divergence: PC users favored a guided, first-person perspective for its focused detail, while VR users benefited significantly from the agency afforded by a third-person perspective with active navigation. 
    After refining each condition based on user feedback---including developing a novel hybrid 1PP/3PP format for PC---Phase 2 compared the two optimized systems (PC vs. VR). 
    Our results show that the immersive VR replay led to significantly better task comprehension and workflow reconstruction accuracy, demonstrating the critical role of embodied agency in understanding complex analytical processes.
} % end of abstract

%% file: sections/01-intro-v2.tex
\firstsection{Introduction}

\maketitle

Immersive analytics is an emerging field that shows significant promise in many applications by leveraging immersive technologies to improve data analysis and decision making~\cite{kraus2022immersive,marriott2018immersive,skarbez2019immersive}.
In general, the primary benefits of this approach stem from the vast spatial canvases~\cite{in2024evaluating,lisle2021sensemaking,yang2020embodied} and the embodied interactions~\cite{huang2023embodied,in2023table,yang2021itvcg,zhu2024compositingvis} that these technologies afford. 
Compared to standard desktop-based analytical workflows, immersive analytics allows analysts to render graphics in a large 3D space around them and interact with data visualizations using intuitive physical movements~\cite{batch2019there,in2025investigating,saffo2023unraveling,yang2025litforager}.

Although these features offer significant advantages for the individual analyst, complex real-world data analysis is now usually conducted by teams in a collaborative manner~\cite{billinghurst2018collaborative,isenberg2011collaborative,sereno2020collaborative,tong2023}. 
Asynchronous collaboration, in particular, has become prevalent due to its flexibility, especially for spatially distributed teams~\cite{heer2007voyagers,lenz2024comparing}. 
In such asynchronous scenarios, a critical workflow is the \textit{handover}, where an incoming analyst must first understand what the previous collaborator completed before they can begin their own work~\cite{walny2019data,zhao2017supporting}.

\added{The challenge for successors in a handover process is to understand insights, intents and analytical history from predecessors. While non-immersive visual analytic tasks often rely on oral communication or provenance tracking tools to facilitate this sensemaking process\cite{zhao2017supporting}, these externalisations fail to capture the predecessor’s spatial focus and embodied interactions, which limits their application in immersive tasks\cite{ens2021grand,lenz2024comparing,zhang2023embodied}. 
To address these challenges, replaying predecessor's session has emerged as a promising method for communicating the analytical process. It helps capture the spatial, temporal and interaction details needed for the successor's task comprehension, and saves the extra burden for the predecessor to explain their analytical history. Past studies have demonstrated the benefits of replay in immersive spatial manipulation tasks, including virtual furniture placement~\cite{wang2019vr} and guided tours~\cite{giovannelli2025investigating}. The design of these replay systems involves critical trade-offs concerning viewing perspective and user agency. Research by Iriye et al.~\cite{iriye2021memories}, for example, has shown that a third-person perspective can enhance spatial memory, while recent work by Giovannelli et al.~\cite{giovannelli2025investigating} indicates that active navigation control enhances engagement and spatial understanding over passive playback.}

The novelty of our work lies in addressing the unique \added{handover challenge} posed by immersive analytics, a domain fundamentally different from prior studies. Previous replay systems have centered on spatial manipulation tasks, such as furniture placement, where the goal is to understand \textit{what object moved where}. In contrast, the core of immersive analytics is \textit{sensemaking}: identifying correlations, filtering datasets, and linking visualizations to form insights~\cite{marriott2018immersive,skarbez2019immersive,tong2025exploring}. The goal of a replay system for immersive analytics is therefore not just to see spatial changes, but to reconstruct the analyst's thought process. This distinction introduces a critical and unaddressed research gap. The information vital to an analytical workflow often requires the detailed, focused view that a first-person perspective provides. Simultaneously, understanding the broader strategy demands the contextual overview of a third-person perspective. This creates a fundamental tension: a third-person view effective for spatial tasks may render crucial data illegible in an IA context, while a first-person view may obscure the overall analytical strategy. Thus, findings from simpler spatial tasks cannot be generalized, and the optimal replay configuration for this cognitively demanding domain remains unknown.

To resolve this critical tension, we designed a two-phase plan to systematically investigate three key design factors (see \autoref{fig:study-overview}): \textit{viewing platform} (PC vs. VR), \textit{perspective} (1PP vs. 3PP), and \textit{navigation control} (active vs. passive). Crucially, our evaluation went beyond simple recall, employing measures such as a \textit{Workflow Reconstruction} task to directly assess comprehension of the sensemaking involved. \textbf{Phase 1} aimed to identify the optimal configuration within each platform by comparing three conditions: 1PP-Passive, 3PP-Passive, and 3PP-Active. The results revealed a clear platform-dependent divergence. While VR users benefited significantly from the agency afforded by 3PP-Active, the findings for PC were more nuanced, with users valuing both the focused detail of 1PP-Passive and the contextual awareness of 3PP-Passive. 

To address this, we developed a novel hybrid 1PP/3PP condition for the PC. Informed by these findings, \textbf{Phase 2} directly compared these optimized formats for PC and VR after enhancing them with user-requested features. Our final results show that the immersive VR replay led to significantly better task comprehension and workflow reconstruction accuracy, providing clear evidence for the benefits of an embodied, active approach for replaying complex analytical processes. Our contributions are:
\begin{itemize}[topsep=1pt, itemsep=0mm, parsep=3pt, leftmargin=9pt]
    \item The design and implementation of replay prototype systems for immersive analytics task handover on both PC and in VR.
    \item \added{Empirical knowledge of the trade-offs between different perspectives (1PP vs. 3PP), navigation control (passive vs. active) and platform (PC vs. VR) for replaying immersive analytic processes in task handover.}
    \item \added{A set of design guidelines for replay systems of immersive analytics tasks.}
\end{itemize}

%% file: sections/02-related-works-v3.tex
\section{Related Work}
\label{sec:related-work}

\textbf{Replay in Extended Reality.}
Replay techniques are widely used in XR to facilitate asynchronous collaboration~\cite{wang2019vr}, tutoring~\cite{li2013ar}, and socialization~\cite{wang2020again}. Unlike screen-based video, replaying immersive sessions requires systematically tracking user inputs, virtual objects, and avatars to reconstruct interactions~\cite{li2013ar,wang2019vr}. Prior systems have demonstrated replay's promise in collaborative contexts like virtual museum tours~\cite{correia2005hypermem} and furniture placement tasks~\cite{wang2019vr}, showing improvements in teamwork and communication clarity. However, these systems are primarily designed for tasks involving simple object transformations (i.e., changes in position, rotation, and scale). This makes them poorly suited for the dynamic nature of immersive analytics, which involves frequent object instantiation, destruction, and complex data-centric interactions. Our work addresses this critical gap.

\vspace{0.1em}
\textbf{Design Factors in Replay Techniques.}
We identify three key design factors from prior work.

\textit{Perspective}: Replay can be presented from a first-person (1PP) or third-person (3PP) perspective. While 3PP is often used in VR to mitigate motion sickness~\cite{ponto2012effective}, the choice involves trade-offs: 3PP can enhance spatial memory~\cite{iriye2021memories}, while 1PP is often associated with a greater sense of immersion~\cite{denisova2015first}. It remains unclear how these perspectives affect the comprehension of complex analytical workflows.

\textit{Navigation}: Viewers can be granted active, user-controlled navigation or guided through a passive, pre-defined camera path~\cite{lu2025ego}. Active navigation can improve spatial learning and engagement~\cite{chrastil2013active, chrastil2012active, giovannelli2025investigating}, but some studies suggest passive navigation can yield comparable results in spatial memory tasks~\cite{gaunet2001active}. These conflicting findings indicate that the effects of navigation are highly task-dependent, motivating our investigation.

\textit{Viewing Platform}: The choice between immersive (VR) and non-immersive (PC) devices has yielded mixed results. Some studies report similar performance in spatial learning tasks, while others find VR can outperform PCs in engagement and retention~\cite{lai2023comparative}. Desktop systems may offer lower immersion but provide stable performance with minimal physiological stress~\cite{kim2012comparison}, highlighting situational advantages of each platform.

\added{
\vspace{0.1em}
\textbf{Immersive Analytics and Task Handover.}
Immersive analytics (IA) leverages VR/AR to enhance data visualization and interaction~\cite{marriott2018immersive}, using embodied interaction to build complex visualizations like scatterplots and parallel coordinate plots in 3D space~\cite{cordeil2019iatk,cordeil2017imaxes}. An important step in collaborative IA tasks is handover, which is process of transferring knowledge and progress between collaborators working asynchronously~\cite{sharma2008sensemaking}. For complex data tasks, a successful handover requires sharing analytical provenance—the historical record of the interactions and reasoning used to reach an insight~\cite{xu2015analytic}. In 2D environment, systems such as VisTrails\cite{freire2012making}  and KTGraph\cite{zhao2017supporting} \ have demonstrated that providing a structured investigation history helps subsequent analysts to recall the critical facts and the predecessor's insights. However, the spatial and embodied nature of immersive analytics makes these handovers more complex, as traditional externalizations like notes or static logs often fail to capture the full context of a predecessor’s interactions and spatial focus\cite{zhang2023embodied}. While recent research into ``embodied provenance'' has begun to address how to track these 3D interactions~\cite{zhang2023embodied}, the effectiveness of using full-session replay as a primary communication tool remains established only for simpler, non-analytical tasks~\cite{correia2005hypermem,wang2019vr}, and the design guidelines to use replay for IA task handover remains underexplored. Our study addresses this gap by investigating how viewing platform, perspective, and navigation control in IA replay impact a user’s ability to comprehend complex, data-rich workflows. }

%% file: sections/03-overview.tex
\section{Project Procedure Overview}
\label{sec:overview}

\begin{figure}
    \centering
    \includegraphics[width=1\linewidth]{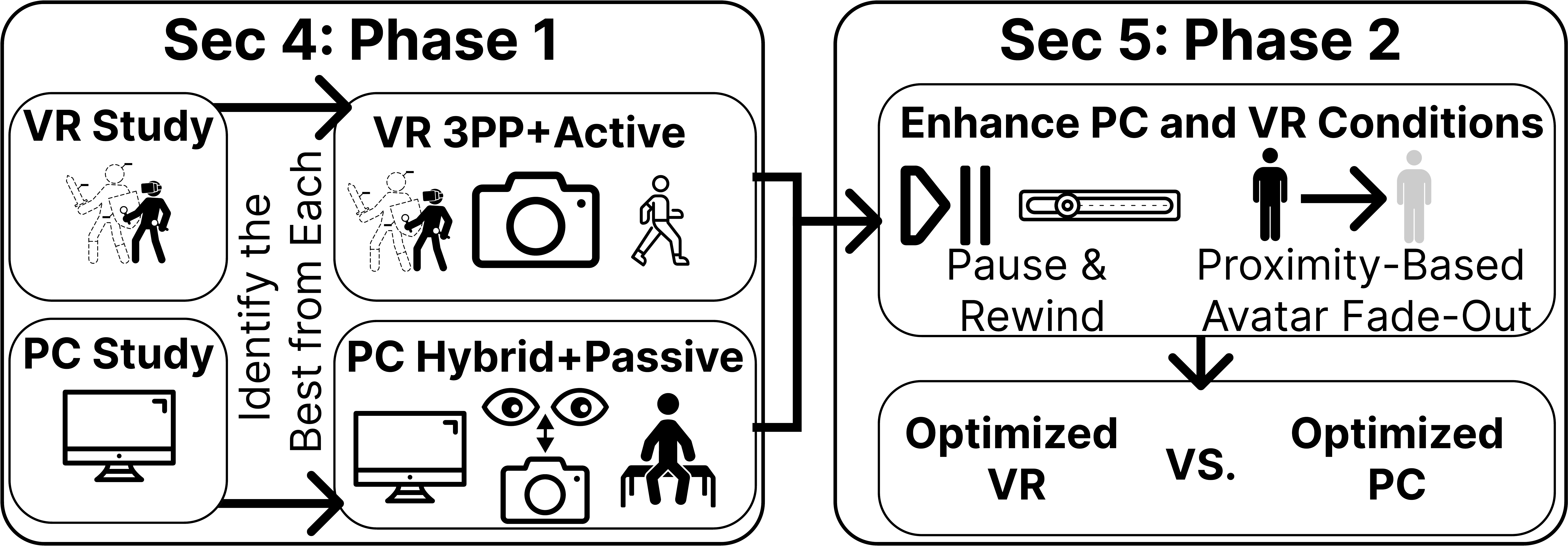}
    \caption{Project Procedure Overview. 
    In Phase 1, we identified the best configuration separately from PC and VR.
    In Phase 2, we first enhanced the conditions by introducing most frequently mentioned features, and then compared the optimized VR and PC conditions.
    \vspace{-0.5em}}
    \label{fig:study-overview}
\end{figure} 

The overarching goal of this research is to systematically investigate the design trade-offs for replaying immersive analytics sessions. 
Informed by prior work in replaying generic VR sessions, we identified three key factors: 
viewing platform (PC vs. VR)~\cite{wang2019vr}, 
perspective (1PP vs. 3PP)~\cite{iriye2021memories}, and 
navigation control (active vs. passive)~\cite{giovannelli2025investigating}. 
A full factorial design would yield eight conditions; however, since a first-person perspective inherently implies passive navigation, the number of valid conditions is reduced to six. This is still too many for a participant to experience in a single session without introducing significant fatigue.
Furthermore, given the absence of prior empirical design guidelines for building immersive analytics replay systems, a single-phase study would risk comparing underdeveloped interfaces. 
We therefore adopted an iterative design process, beginning with basic prototypes, gathering feedback through evaluation, and refining the prototypes for a final comparative study. 
To ensure the feasibility of this approach and to generate more polished designs, we structured our systematic evaluation into a two-phase methodology (see \autoref{fig:study-overview}).

In \textbf{Phase 1}, our objective was to determine the optimal replay configuration for each platform (PC and VR) independently. 
To achieve this, we conducted two separate within-subject user studies, with each study focusing on one platform. Within each platform, we investigated the effects of perspective and navigation control on the replay experience across three study conditions: 
(1) 1PP with passive navigation, 
(2) 3PP with passive navigation, and 
(3) 3PP with active navigation.

\textbf{In Phase 2}, we first analyzed the results and qualitative feedback from Phase 1 to refine our prototypes. 
Based on this analysis, we implemented two additional features that were frequently requested by participants. We then conducted a comparative evaluation between the two improved designs. 
By incorporating the newly implemented user-centered features, this final study aimed not only to evaluate the trade-offs between viewing platforms (PC vs. VR) but also to provide more ecologically valid design guidelines and implications.

\added{The proposed study procedure was approved by the Institutional Review Board (IRB) in our university before it was conducted.}

%% file: sections/04-study1.tex
\section{Phase 1: Identifying the Optimal Replay Configurations within PC and VR}
\label{sec:study-1}

The first phase was designed to systematically evaluate the impact of perspective and navigation control on the effectiveness of replaying immersive analytics sessions. 
The primary goal of Study 1 was twofold: (1) to empirically determine the optimal replay configuration within both PC and VR platforms independently, and (2) to gather evidence-based insights that would inform the development of improved, user-centric replay systems for each medium.

\begin{figure}
    \centering
    \includegraphics[width=1\linewidth]{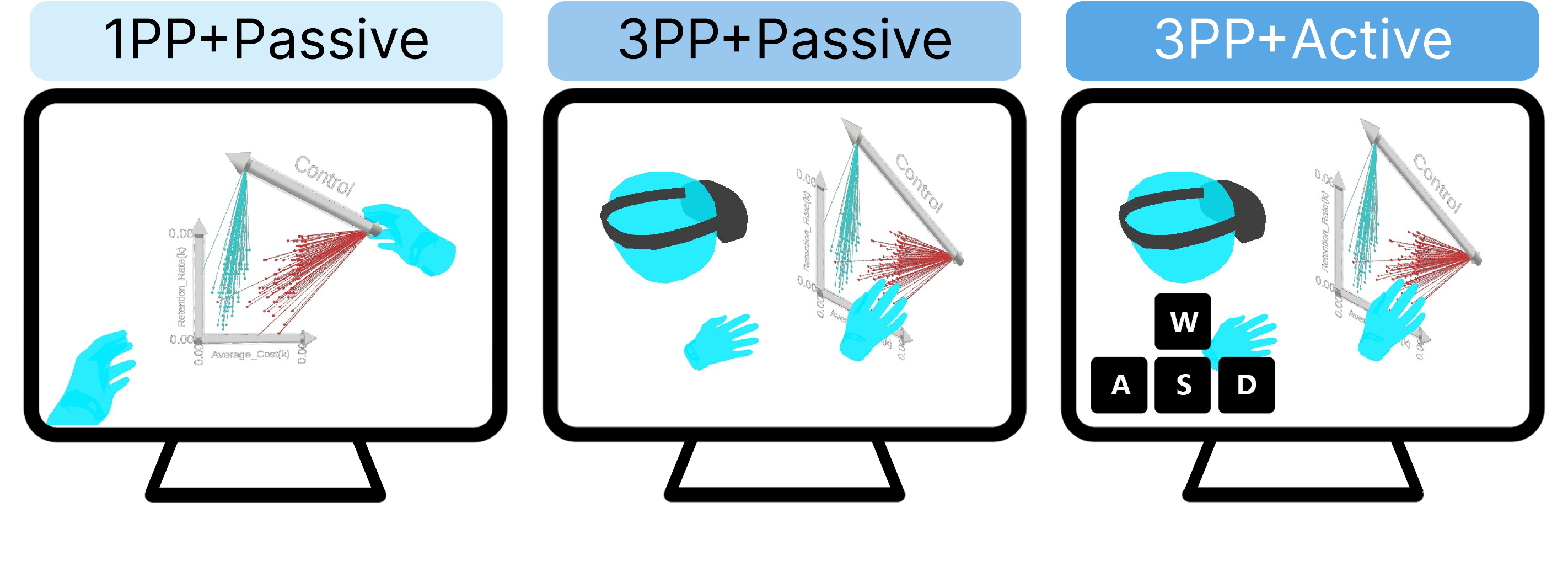}\vspace{-1em}
    \caption{The three PC conditions evaluated in Phase 1.
    \textbf{1PP-Passive}: A first-person view from the original analyst's perspective with no user interaction.
    \textbf{3PP-Passive}: A third-person view that follows the analyst's avatar along a pre-defined spatial trajectory.
    \textbf{3PP-Active}: A third-person view that permits viewer-initiated navigation through mouse and keyboard, like some video games.\vspace{1em}}
    \label{fig:study-1-pc}
\end{figure}

\subsection{Study Design and Conditions}
\begin{figure}
    \centering
    \includegraphics[width=1\linewidth]{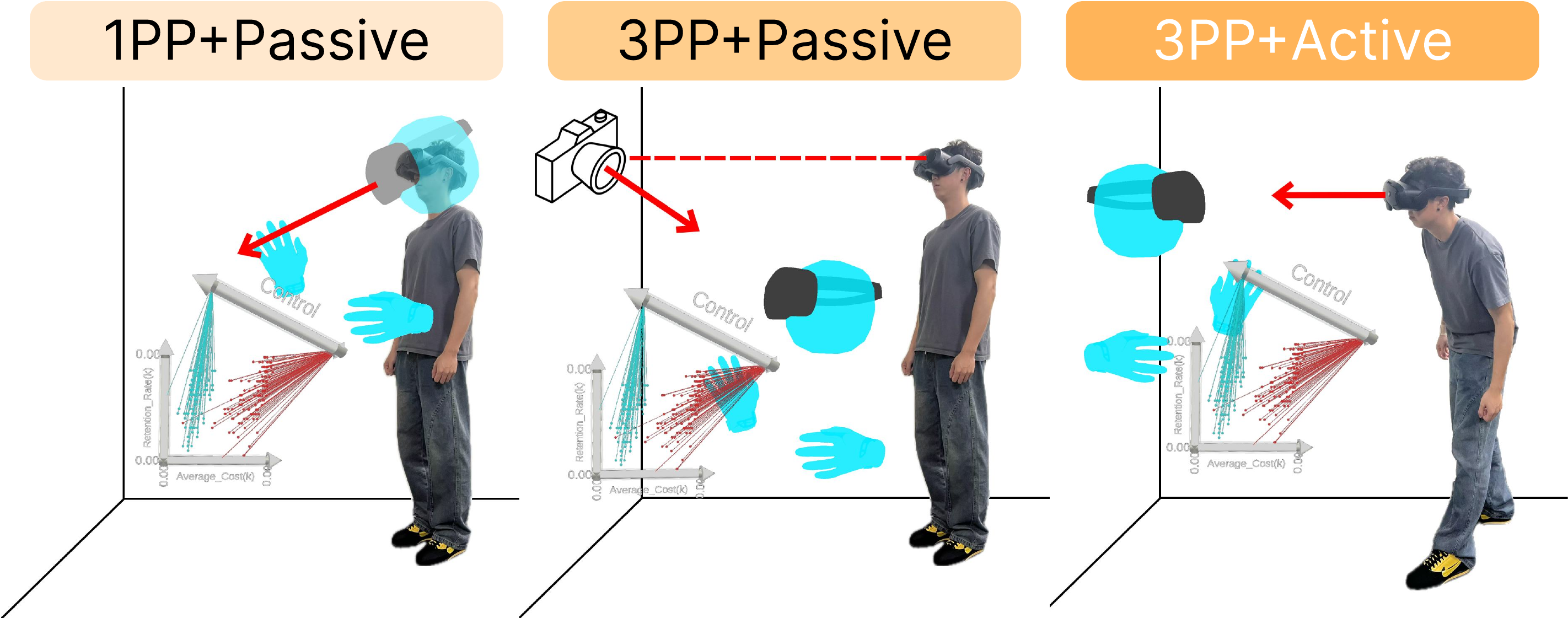}
    \vspace{-0.5em}
    \caption{The three VR conditions evaluated in Phase 1.
    \textbf{1PP-Passive}: A first-person view from the original analyst's perspective with no user interaction.
    \textbf{3PP-Passive}: A third-person view that follows the analyst's avatar along a pre-defined spatial trajectory.
    \textbf{3PP-Active}: A third-person view that permits viewer-initiated navigation through body movements.\vspace{-0.5em}}
    \label{fig:study-1-vr}
\end{figure}

Study 1 employed a within-subjects experimental design for each of the two viewing platforms. 
Each participant experienced three distinct replay conditions, see \autoref{fig:study-1-pc} for illustrations of PC conditions and \autoref{fig:study-1-vr} for VR conditions.
The order of these conditions, along with the three experimental tasks, was fully counterbalanced using a Latin Square design to mitigate learning and fatigue effects. 
The three conditions experienced by each participant were: 1PP+Passive, 3PP+Passive, and 3PP+Active, as described below. 
% The total six conditions in study 1 were proposed to systematically explore the design space.

\textbf{1PP+Passive}: 
In this condition, the participant views the session from the original analyst's viewpoint.
\textbf{On the PC}, this was presented as a standard, pre-recorded first-person video.
\textbf{In VR}, early pilot testing revealed that directly mapping the raw, recorded head movements to the user's HMD induced significant motion-sickness due to the mismatch between the user's physical head motion and the virtual motion. 
To mitigate this, we implemented a system analogous to a 360-degree video. Participants were seated and could freely move their heads to look around, but their virtual position in the environment was determined by the analyst's recorded path. 
To reduce motion sickness from erratic head movements, we smoothed the recorded head position and rotation data using a one-second moving average filter, resulting in a more stable and comfortable viewing experience.

\textbf{3PP+Passive}:
In this condition, the participant observes the original analyst, represented by a virtual avatar, from an external viewpoint.
On both PC and in VR, the camera followed a carefully choreographed, pre-recorded path designed to capture the analyst's activities and their interactions with the data visualizations. 
As no mature techniques currently exist for automatically generating an optimal camera path for such tasks, we manually recorded the camera path for each task replay. 
This path was standardized across all participants and was designed to focus on the most salient information, including the specific visualizations being manipulated and the avatar's body language and gaze direction.

In the VR version, similar to the 1PP+Passive VR condition, the participant's viewpoint was fixed to the moving camera path, but they retained the freedom to move their head to observe the scene from that viewpoint, analogous to being on a guided tour track.

\textbf{3PP+Active}:
This condition granted participants full control over their movement through the recorded scene while observing the analyst's avatar.

In VR, this was inherently supported, allowing participants to freely and physically walk around the virtual environment to explore the scene from any position or angle.

On PC, we used a control scheme common in video games. 
Participants used the W, A, S, and D keys for planar translation (forward, left, backward, and right)~\cite{enriquez2024evaluating}. 
To accommodate vertical movement, which pilot users indicated was desirable for getting better views of 3D visualizations, two additional keys (Q and E) were mapped for moving up and down. Rotational control of the camera (yaw and pitch) was mapped to mouse movement.

% \subsection{System Implementation.}
The replay system was developed in the Unity3D game engine. For the immersive analytics environment, we adapted ImAxes~\cite{cordeil2017imaxes}, an open-source Unity3D-based library that supports the embodied creation and interaction with visualizations in VR. The replay functions were implemented through frame-based scene logging.

\subsection{Datasets and Tasks}
\added{
In each condition, the overall task for participants is first viewing a replay of an IA session, and then answer questions on task comprehension and reconstruct the wrokflow. }To ensure our findings were grounded in realistic scenarios, we selected four widely-used datasets to produce those replay sessions, including the Starbucks nutrition dataset, the US Colleges dataset, a wine quality dataset, and the Spotify song attributes dataset. The Spotify dataset was used for the initial training session, while the other three were used for the main experimental tasks. 
To control task difficulty and ensure comparability across the within-subject conditions, each dataset was manually curated to contain 10-12 variables and approximately 200 data points.
All replay sessions were designed as structured exploratory analysis process. 
Each replay depicted an analyst investigating a central theme (e.g., identifying chemical properties that influence wine quality) while also exploring other emergent questions. 
The analyst's actions were adapted from a visual analytics task paradigm~\cite{shneiderman2003eyes}, including  Construct, Discard, Categorize, Filter, Explore (Details-on-demand) and Identify. 
The visualizations created included 2D/3D scatterplots, linked visualizations, and parallel coordinate plots from the ImAxes system~\cite{cordeil2017imaxes}. 
By keeping the number of analytical steps and visualizations constructed consistent across the three sessions, we maintained a comparable level of complexity.

The purpose of Phase 1 is to compare different perspective and navigation control settings. To avoid confounding factor of unnecessary interactions, we did not provide pausing or rewinding the replay. \added{Participants watched the replay sessions uninterrupted from the beginning to the end.} Each session was about six minutes.

\subsection{Participants}
A total of 24 participants were recruited for Study 1 from the local university community through mailing lists and campus announcements. Participants were required to be over 18 years of age with normal or corrected-to-normal vision and no prior history of conditions that could be exacerbated by VR use, such as severe motion sickness. The recruited participants were 13 males and 11 females, aged between 19 and 29 years (\textit{M} = 24.1, \textit{SD} = 2.3). We also collected their  self-reported previous experience with VR (\textit{M} = 2.52, \textit{S}D = 1.05) and data visualisation(\textit{M} = 3.68, \textit{SD} = 0.90).They were randomly and evenly assigned to one of two independent groups: the PC Group (\textit{N} = 12) or the VR Group (\textit{N} = 12).

\subsection{Procedure}
Each experimental session lasted approximately 85 minutes and followed a standardized procedure. 

After providing informed consent and completing demographic and pre-study questionnaires, participants underwent a 10-minute training session, which was later repeated before each condition correspondingly. This session was designed to familiarize them with the replay interface for their assigned medium (PC or VR), the perspective and navigation control, and the nature of the immersive analytics tasks, using the pre-prepared Spotify music dataset. 

Following the training, participants proceeded through the three counterbalanced experimental conditions. Before each condition, they were given a printed document providing the context for the specific analysis task (i.e., the College, Wine, or Starbucks dataset). 
They then watched the corresponding six-minute replay of the analyst's session. They were instructed to look for answers to the \textit{Comprehension Questionnaire}. 
In VR, the questions were shown on a wrist-mounted UI panel. The user can look at the questions when they feel necessary. 
On the PC, the questions were printed on a paper for the same purpose. Immediately after each replay, participants completed the two primary task performance measures: the \textit{Comprehension Questionnaire} and the \textit{Workflow Reconstruction} task. Afterward, they filled out the NASA-TLX and a user satisfaction questionnaire. A mandatory three-minute break was provided between each condition to minimize fatigue.

Upon completing all three conditions, participants were asked to rank their preference for each format. The session concluded with a ten-minute semi-structured interview to gather in-depth qualitative feedback on their experience, the challenges they faced, and the perceived benefits of each replay format.

\subsection{Measures}
Our evaluation focused on holistically assessing each participant's understanding of both (a) the data-related insights and (b) the analytical workflow used to reach those insights. We employed a mix of quantitative and qualitative measures.

\textbf{Task Performance:} We apply \textbf{\textit{Comprehension Questionnaire:}} A five-item questionnaire designed to test understanding of the session. The questions assessed recall of basic visual details, comprehension of simple data insights and understanding of the analyst's intent.
\textbf{\textit{Workflow Reconstruction:}} Participants were given a pen and paper and asked to reconstruct the entire analytical workflow they had just observed. They were instructed to use the pre-defined action types to create a flowchart representing the analyst's process.

\textbf{Subjective Measures:} After each condition, participants completed standard psychometric questionnaires:
\textit{Cognitive Load:} Measured using the NASA-TLX questionnaire.
\textit{Satisfaction:} Measured using a standard user satisfaction rating scale.

\textbf{Qualitative Feedback:} At the end of the study, a semi-structured interview was conducted to gather in-depth qualitative feedback on participants' experiences, preferences, and the reasoning behind their performance.

We did not compare the completion time, as the completion time was determined by the sessions rather than the user performance. All sessions also share similar durations.

\subsection{Hypotheses}
Our hypotheses for Phase 1 were formulated based on previous literature \cite{lu2025ego,gaunet2001active,iriye2021memories} to address how perspective and navigation control impact a user's task understanding, cognitive load, and user satisfaction in a replayed immersive analytics session. 
These hypotheses were tested separately within both the PC and VR groups.

% \textbf{Perspective} (1PP vs. 3PP):
$H(Pers\cdot Perf)$: We hypothesized that the 3PP-Passive condition would lead to better task performance (higher comprehension and workflow reconstruction scores) compared to the 1PP-Passive condition, as the exocentric view provides greater situational awareness.

$H(Pers\cdot Load)$: We hypothesized that the 3PP-Passive condition would result in a lower cognitive load and higher satisfaction compared to the 1PP-Passive condition, as it reduces the mental effort for spatial alignment and perspective-taking from a potentially disorienting first-person viewpoint. 

% \textbf{Navigation Control} (Passive vs. Active):
$H(Nav\cdot Perf)$: We hypothesized that the 3PP-Active condition would result in better task performance compared to the 3PP-Passive condition. We predicted that user-directed navigation would allow participants to seek information more effectively and clarify spatial relationships on their own terms.

$H(Nav\cdot Load)$: We hypothesized that while active navigation might increase physical workload, it would result in lower cognitive load and higher user satisfaction, benefited from the agency to control one's viewpoint.

\subsection{Results}
\begin{figure}[t]
    \centering
    \includegraphics[width=1\linewidth]{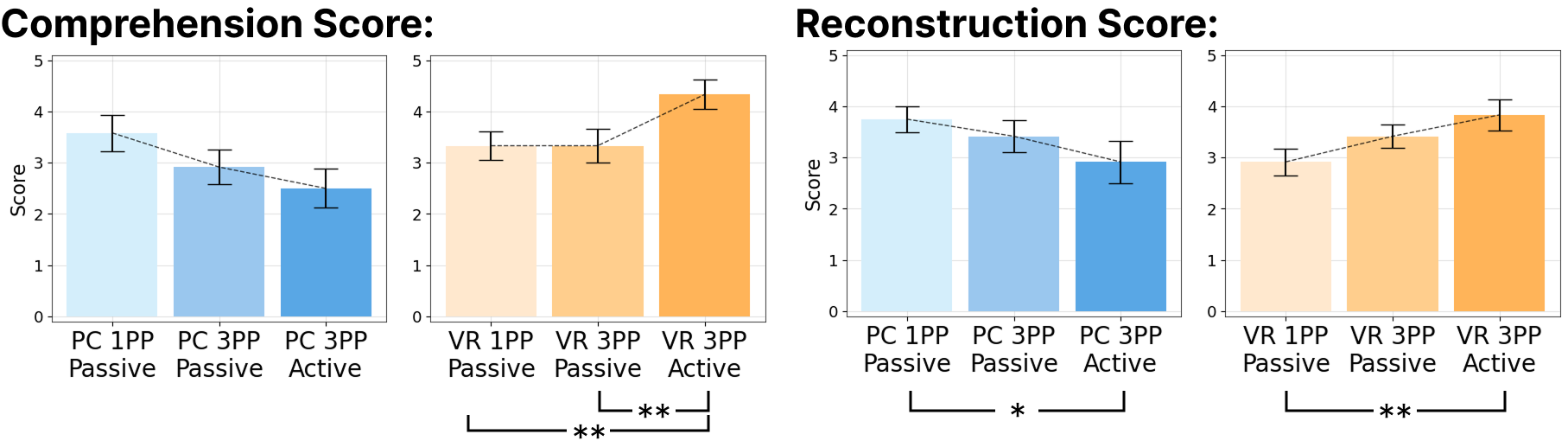}
    \caption{Average Comprehension and Reconstruction Scores in PC and VR Conditions. * indicates borderline significance ($.05< p<.1$), ** indicate significance ($p<.05$)}
    \label{fig:fig1}

    \vspace{\baselineskip} 
    \vspace{\baselineskip}

    \includegraphics[width=1\linewidth]{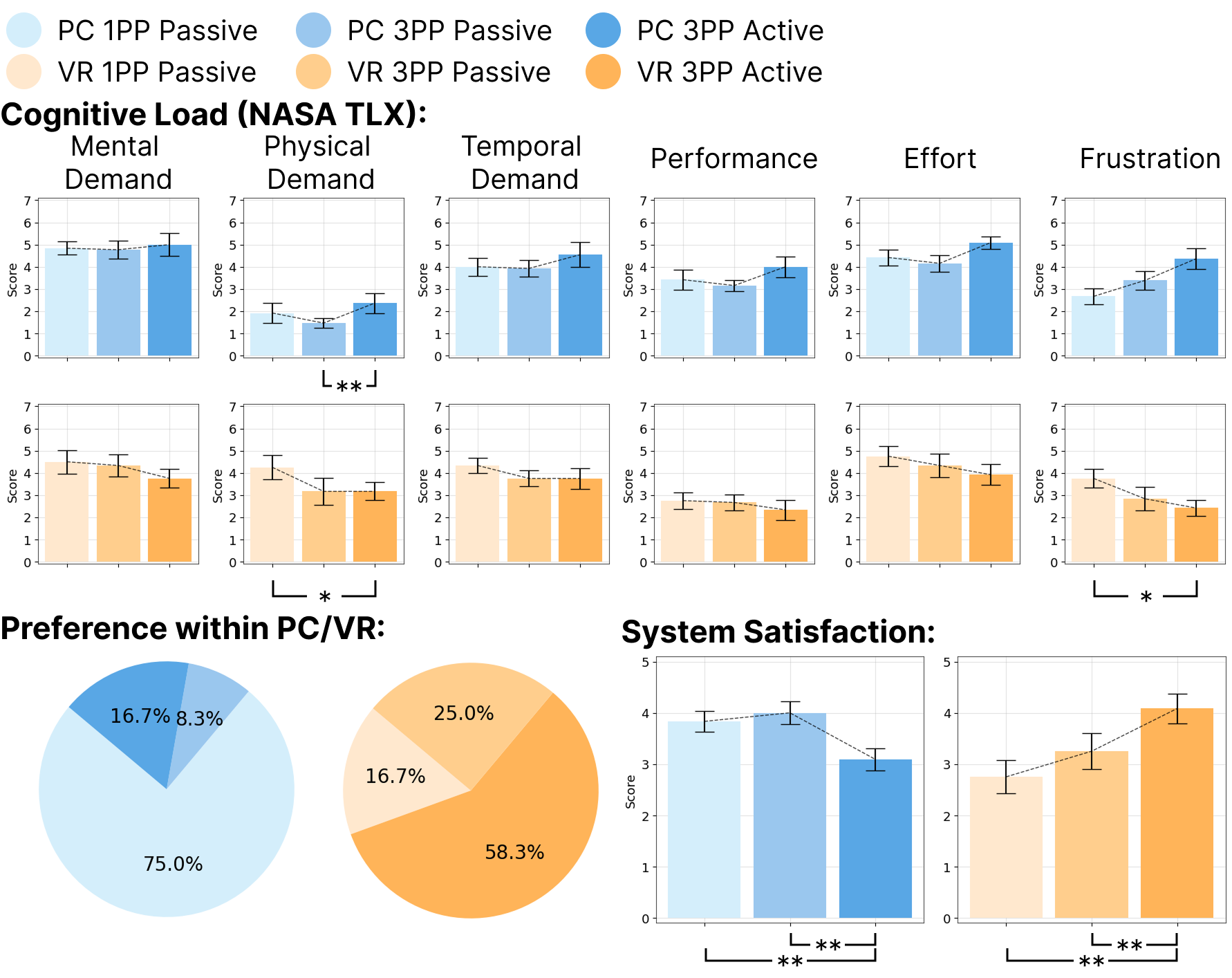}
    \caption{Subjective ratings for PC and VR conditions: Cognitive Load, Preference and Satisfaction}
    \label{fig:fig2}
\end{figure}

This section presents the statistical analysis of the quantitative data from Phase 1. 
Our analysis aimed to identify the optimal replay configurations for both the PC and VR media platforms. 
For all tests, we report $p<0.05$ as statistically significant, and $p<0.1$ as marginally significant. 
Given that Shapiro-Wilk tests indicated that the data did not consistently meet the assumptions of normality, we employed non-parametric methods for our analysis. 
The Friedman test was used to detect overall differences among the three within-subjects conditions. 
When a significant effect was found, post-hoc pairwise comparisons were conducted using the Wilcoxon signed-rank test with a Bonferroni correction for multiple comparisons. Task performance results and subjective measure results are shown in \autoref{fig:fig1} and \autoref{fig:fig2} respectively.

\subsubsection{VR Group}
For the 12 participants in the VR group, analysis of task performance and preference revealed a clear advantage for the active navigation condition.

\textbf{Task Performance:}
A Friedman test showed a statistically significant effect of the replay condition on the workflow reconstruction scores ($\chi^2(2)=13.56, p=.001$).
Post-hoc tests revealed that participants in the 3PP-Free Navigation condition achieved significantly higher reconstruction scores than those in both the 1PP-Passive ($Z=0.0, p=.015$) and 3PP-Passive ($Z=0.0, p=.010$) conditions.

Similarly, a significant effect was found for the comprehension question scores ($\chi^2(2)=9.77, p=.008$). 
Pairwise comparisons again showed that the 3PP-Free Navigation condition significantly outperformed both the 1PP-Passive ($\mathit{Z} = 4.0,\ \mathit{p} = .038$) and 3PP-Passive (\textit{Z} = 0.0, \textit{p} = .047) conditions. These results support $H(Nav\cdot Perf)$ and $H(Per\cdot Perf)$, indicating that for VR-based replays, providing users with agency to freely navigate the scene leads to a significantly better understanding of both the analytical workflow and the data insights.

\textbf{Subjective Ratings:}
User preference data strongly corroborated the performance results. A Friedman test on the preference rankings was statistically significant ($\chi^2(2)=6.20, p=.045$), with 3PP-Free Navigation being the most preferred condition. Analysis of the NASA-TLX data also showed significant differences in Frustration ($\chi^2(2)=6.00, p=.049$) and Satisfaction ($\chi^2(2)=9.05, p=.011$).Post-hoc tests revealed that participants were significantly less frustrated (\textit{Z} = 12.5, \textit{p} = .0065)  and more satisfied(\textit{Z} = 9.0, \textit{p} = .031) in the 3PP-Free Navigation condition compared to the 1PP-Passive condition.

\textbf{Feedbacks: }
Qualitative feedback from interviews strongly supported the quantitative preference for the 3PP-Active condition, revealing three key themes: \textit{\textbf{Agency in Free Navigation Was Critical for Comprehension.}}
The preference for 3PP-Active was driven by the sense of control it afforded. Eight participants highlighted that freedom of movement allowed them to investigate the scene from optimal angles and revisit points of interest. As P12 summarized, \textit{``Free navigation felt most natural... it offered the most flexible angle.''} This ability to \textit{``go around and double check''} (P1) was directly linked to better performance and lower cognitive load, as it empowered users to resolve their own confusion. \textit{\textbf{Passive Viewpoints Caused Disorientation and Frustration.}}
Conversely, passive conditions were criticized for being restrictive and uncomfortable. Seven participants described the 1PP condition as dizzying and unnatural, with P2 noting, \textit{``First person was more dizzy, the mental strain was higher.''} A key issue was the lack of agency, which made p7 \textit{``feel a loss of control''}. While the 3PP-Passive condition was an improvement, its pre-recorded camera path was still problematic; some found its movements disorienting (P10) or the viewpoint too distant from important details (P3).
\textit{\textbf{Visual Occlusion Was a Major Obstacle.}}
A recurring complaint across both third-person views was visual occlusion by the analyst's avatar. Six participants mentioned that the avatar's head would occasionally block critical information. As P4 stated, the avatar ``helped in understanding the intention, except for blocking the view.''

\subsubsection{PC Group}
In contrast to the VR group, the results for the 12 participants in the PC group were less conclusive regarding task performance but still revealed clear trends in user experience and preference.

\textbf{Task Performance: }Friedman tests on the performance data for the PC group showed no statistically significant differences across the three conditions for either workflow reconstruction ($\chi^2(2)=3.62, p=.016$) or comprehension question scores ($\chi^2(2)=3.35, p=.019$). 
However, pairwise comparisons revealed marginal significance exists between 1PP Passive and 3PP Active conditions (\textit{Z} = 3.0, \textit{p} = .09) in workflow reconstruction, which does not support $H(Nav\cdot Perf)$ or $H(Nav\cdot Perf)$. As an extension, we conducted between-group analysis in corresponding PC and VR conditions. While there were no significant performance differences between passive viewing conditions, the VR group achieved significantly higher comprehension scores than the PC group (\textit{U} = 20.50, \textit{p} =.0024) in the 3PP-Active condition, highlighting the platform-dependent impact.

\textbf{Subjective Ratings: }While performance was comparable across conditions, subjective ratings revealed notable differences. A Friedman test on preference rankings was significant ($\chi^2(2)=8.00, p=.018$), with the 1PP-Passive video format being the most preferred. The 3PP-Free Navigation condition was rated as the least preferred. 
The NASA-TLX results showed a borderline significant difference in Physical Demand ($\chi^2(2)=5.84, p=.005$), with post-hoc tests indicating that the 3PP-Free Navigation condition was perceived as significantly more physically demanding than the 3PP-Passive condition (\textit{p} = .034). There was also a borderline significant difference in Satisfaction ($\chi^2(2)=4.75, p=.093$), with pairwise comparisons showing that both the 1PP-Passive and 3PP-Passive conditions were significantly more satisfying than the 3PP-Free Navigation condition (\textit{p} = .039 and \textit{p} = .045, respectively). This suggests that on a desktop, the additional effort required for manual navigation did not translate into a better user experience and was in fact detrimental to satisfaction. The results reject both $H(Nav\cdot Pref)$ and $H(Per\cdot Pref)$.

\textbf{Qualitative Feedback:} The interviews with the 12 participants in the PC group revealed a strong preference for passive, guided experiences. The key feedback includes: \textbf{\textit{Preference for Guided, Focused Views.}} The most prevalent theme was a preference for the 1PP-Passive video condition, which 9 out of 12 participants ranked as their top choice. The primary reason was that it minimized distractions and cognitive overhead. In contrast, the third-person views had \textit{``more visual contents that are not related to data insight.''} Participants felt direct perspective made it easier to follow the analyst's focus.  \textbf{Free Navigation on PC Increases Cognitive Load.} In opposition to the VR group, 3PP-Free Navigation was the least preferred. Eight participants ranked it last, citing the difficulty of control and the increased mental effort required. P23 found that having to manually control the camera \textit{``consumes attention and makes you skip important information.''} P22 elaborated on this, stating that \textit{``I'd rather focus on data, instead of paying attention to control it on PC.''} This difficulty was compounded by unfamiliar controls (P23) and the challenge of simultaneously tracking the avatar and the complex 3D visualizations (P18). \textbf{Avatar Occlusion and Distraction.} Similar to the VR group, visual occlusion by the avatar was a frequently mentioned issue in the 3PP and Free Navigation conditions. However, on the PC, some participants also perceived the avatar itself as a distraction(P18, P21, P24). This indicates that without a strong sense of co-presence, the on-screen avatar could be seen as unnecessary visual element.

\subsection{Discussion}
The results of Study 1 provide key insights into how the choice of perspective and navigation control impacts the user experience differently on PC and VR platforms. The most interesting finding is the clear divergence in user preference and performance between the two media groups, indicating that the optimal design for an immersive analytics replay system is highly platform-dependent.

\textbf{The Effect of Navigation Control: Agency in VR and Cognitive Load on PC.}
Our study reveals a fundamental difference in how users perceive and benefit from navigation control in VR versus on a PC. For the \textbf{VR} group, providing active navigation (3PP-Free) led to a statistically significant improvement across all measures: task performance, user preference, and satisfaction. This finding aligns with prior work on spatial learning, which posits that active, embodied exploration leads to the formation of more robust cognitive maps and enhances comprehension \cite{chrastil2015active}. In our study, the embodied nature of VR made physical navigation feel intuitive. As participants noted in interviews, the freedom to reposition themselves to get a better vantage point or ``go and double check'' details was crucial for identifying data from different dimension, which were significant barriers in the passive conditions. This user agency appears to be a key factor in reducing frustration and improving overall task understanding.

Conversely, the \textbf{PC} group showed the opposite trend. The 3PP-Free Navigation condition was the least preferred, rated as significantly less satisfying, and was perceived as more physically demanding than the passive conditions. Crucially, this increased effort did not translate to better performance, as there was no significant improvement in task scores. This finding is consistent with Cognitive Load Theory \cite{sweller1991evidence}, which suggests that mentally demanding tasks unrelated to the primary learning objective can hinder performance. On a 2D screen, navigating a 3D space with a keyboard and mouse is an abstract, learned skill that imposes significant cognitive load. As participant P22 mentioned, it is “better to focus on data, instead of paying attention to control it on PC.” We also observed that some users in this condition would navigate to a single, preferred viewpoint and then remain stationary for the majority of the replay, effectively opting out of continuous navigation. For desktop users, the mental effort of navigation competed with the primary task of understanding the analysis, making guided, passive conditions a more efficient choice.

\textbf{The Effect of Perspective: Situational Awareness vs. Guided Focus.}
The choice of perspective also yielded different results across the platforms, highlighting a trade-off between the situational awareness of a third-person view and the direct focus of a first-person view.

In \textbf{VR}, the \textbf{1PP} condition was consistently rated as the least preferred and was a significant source of discomfort for many participants, confirming that replaying raw, first-person head movements can be a disorienting experience. The \textbf{3PP} view, which provided a more stable, external perspective, was preferred, as it allowed users to understand the analyst's actions within the broader context of the virtual environment.

On the \textbf{PC}, however, the \textbf{1PP}-Passive video was the most preferred format. Participants found this view to be the most efficient and least distracting, as it directed their attention precisely where the original analyst was looking, with no avatar to cause occlusion. On a limited screen space, the 3PP view could place key visuals further away, and the avatar itself was occasionally occluding with visualisations.  However, our qualitative feedback also indicated that while the 1PP view was better for seeing details, the 3PP view was still useful for gaining a contextual overview, especially when more than one visualisation were explored. This suggests that for PC users, both a clear, ego-centric focus for detail and a broader, exocentric view for context are important.

\textbf{Deriving the Optimal Configurations for Study 2}

Our findings from Study 1 demonstrate that a universal approach to designing replay systems is suboptimal. Therefore, two different formats were chosen for Study 2 configurations. 

For \textbf{VR}, the data unanimously points to 3PP-Free Navigation as the superior format. It yielded the best task performance, the highest user satisfaction, and was the most preferred condition. The benefits of embodied agency in VR are clear.

For the \textbf{PC}, the optimal choice is less straightforward. The 1PP-Passive condition was preferred for its focused view, while the 3PP-Passive view was valued for its contextual awareness. The two conditions did not produce a salient difference in task performance. Recognizing this trade-off, and in response to user feedback (P17) suggesting a desire for both, we chose to develop an improved, hybrid 1PP/3PP-Passive condition for Study 2. This design aims to synthesize the benefits of both perspectives, allowing users to switch freely between the focused view of 1PP for detailed analysis and the contextual overview of 3PP for understanding the broader workflow.

Furthermore, based on consistent feedback from both groups, we identified two additional features to be implemented in the systems for Study 2. First, to address the frequently mentioned issue of avatar occlusion, a \textbf{proximity-based fade-out} feature will be added, making the avatar more transparent as the user gets closer. Second, to address the high temporal demand reported in the NASA-TLX scores and make the settings more applicable for complex IA tasks, \textbf{pause and rewind} capabilities will be integrated. 

%% file: sections/05-study2.tex
\section{Phase 2: Comparing Optimized PC and VR replay Systems}

The second phase of the user study directly compared the optimized replay configurations for PC and VR that were derived from the findings of Study 1. \added{By comparing these two formats, we aim to compare the optimized VR and PC replay system in terms of task performance, cognitive load, and user satisfaction, and produce empirical insights on the design of IA replay systems.}

\begin{figure}
    \centering
    \includegraphics[width=1\linewidth]{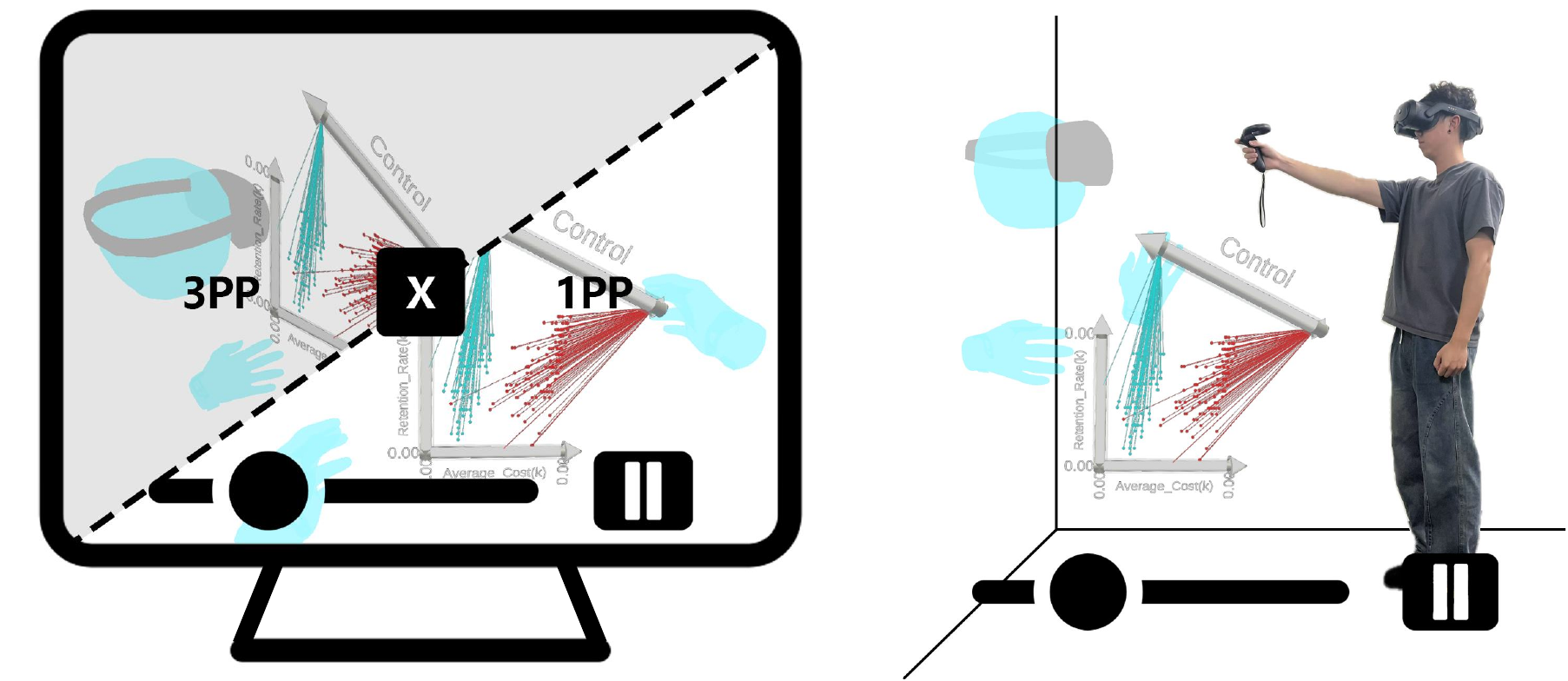}\vspace{-0.5em}
    \caption{The two conditions evaluated in Phase 2.
    \textbf{PC Hybrid (1PP/3PP-Passive)}: A guided experience where the viewer cannot navigate freely but can instantly switch between 1PP and 3PP.
    \textbf{VR (3PP-Active)}: An embodied experience where the viewer observes from a 3PP and has full navigational control via physical movement. 
    Both conditions were enhanced with user-requested features from Phase 1, including full playback controls and a proximity-based avatar fade-out to mitigate visual occlusion. 
    \vspace{-0.5em}}
    \label{fig:study-2-conditions}
\end{figure} 

\subsection{Conditions and System Improvements}
Based on the results and qualitative feedback from Phase 1, we developed two optimized replay conditions for this second phase. \textbf{Optimized VR Condition} was 3PP+Active, while \textbf{Optimized PC Condition} featured a novel hybrid 1PP/3PP+Passive format. 
This design was created to synthesize the benefits of both passive perspectives from Study 1, addressing the trade-off between the focused, detailed view of 1PP and the contextual, situational awareness of 3PP. 
Users could instantly switch between the two perspectives at any time by pressing a key.

% \textbf{System Improvements:}
Additionally, two new features were added for both conditions:

\textbf{Pause and Rewind:} Participants were given full control over the playback timeline. 
On PC, a standard slider and pause button were provided. 
In VR, when the user presses the controller trigger, a timeline slider appears in front of them. By holding the trigger and moving the controller horizontally, they can drag the slider to the left or right and set a new time point. This feature was added to represent real-world use cases by giving users the agency to revisit specific moments and reduce the temporal demand of the task.

\textbf{Proximity-Based Avatar Fade-Out:} To address the prevalent issue of avatar occlusion, a dynamic fade-out feature was implemented. The analyst's avatar would automatically become more transparent as the user's viewpoint moved closer to it within 1.5 meters threshold, ensuring that the user's view of the data visualizations was clear.

\subsection{Tasks and Measures} 
The tasks for Phase 2 used the Starbucks and College datasets from the first study, with the Spotify dataset again used for training. 
To better leverage the new playback controls and to increase the task's analytical depth, the replayed sessions were redesigned to include more analytical steps and visualizations. Each session last about 30 minutes.
The measures were also refined to capture quantified task performance and adapt to the new tasks: 

\textbf{\textit{Comprehension Questionnaire:}} 
The questionnaire was expanded to seven items, including two questions on \textit{visual details} (e.g. what did blue and red represent in the visualisations respectively?), 
two on the \textit{analyst's intention} (e.g. how did the author group the variables in the first step?), and 
three on \textit{data-related insights} (e.g. what was the relationship between retention rate and admission rate?), with one question designed to test a \textit{deeper understanding across multiple analytical steps} (e.g. Compare the difference of data distribution in average cost and graduate earnings).

\textbf{\textit{Spatial Recall:}} 
A new measure was introduced to specifically test spatial memory. After each replay, participants were asked to use a pen and paper to draw the relative position of a chosen visualization from the session.

\textbf{\textit{Interaction Logging:}} We quantitatively logged user interactions, including total task completion time, the number and timing of pauses and rewinds, the number of perspective switches in the PC condition, and the frequency and duration of question panel views.

Other measures, including workflow reconstruction, NASA-TLX, and satisfaction ratings, remained the same as in Phase 1.

\subsection{Participants and Study Design}
A new group of 12 participants, recruited using the same criteria as Phase 1, took part in the second study. The recruited participants were 7 males and 5 females, aged between 18 and 25 years (\textit{M} = 22.5, \textit{SD} = 1.1). We also collected their  self-reported previous experience with VR (\textit{M} = 3.08, \textit{S}D = 1.08) and data visualization (\textit{M} = 3.5, \textit{SD} = 1.09).
The study employed a within-subjects design, where each participant experienced both of the refined replay conditions on PC and in VR. The order of the two conditions as well as the two experimental tasks were counterbalanced using a Latin Square design.

\subsection{Procedure}
Each session lasted approximately 70 minutes. 
The procedure was similar to that of Phase 1, with participants first completing a training session for one of the conditions. 
Before each replay, they were given the task context. The comprehension questions were made available to the participants during the replay. On the PC, the questions could be toggled on and off with a keypress. 
In VR, the questions were displayed on a wrist-mounted UI panel. 
After completing the first condition and its associated measures, participants took a break before proceeding to the second training and experimental session with the other condition. The study concluded with a final preference ranking and a semi-structured interview.

\subsection{Hypotheses}
Our hypotheses for Study 2 focused on the comparison between the two optimized media platforms.

\textbf{$H(Perf)$:} We hypothesized that the VR condition would lead to better task performance (shorter task completion time, higher comprehension, workflow, and spatial recall scores) than the PC condition, due to the benefits of embodied agency and spatial immersion.

\textbf{$H(Load)$:} We hypothesized that the VR condition would result in a higher physical workload and cognitive workload compared to the PC condition.

\textbf{$H(Pref)$:} We hypothesized that the VR condition would be rated as more satisfying and would be the preferred format due to its immersive and engaging nature.

\subsection{Results}
This section presents the statistical analysis comparing the optimized PC and VR replay formats from Phase 2. 
Our analysis aimed to evaluate differences in task performance, cognitive load, and user satisfaction.
We used the same statistical methods as of Phase 1.
% A significance threshold of .10 was used to determine statistical significance. Given normality could not be assumed, we primarily non-parametric methods. The Wilcoxon signed-rank test was used to compare performance metrics (comprehension, workflow, and spatial recall) and user satisfaction scores, as well as the NASA-TLX subscales.

\begin{figure}[t]
    \centering
    \includegraphics[width=1\linewidth]{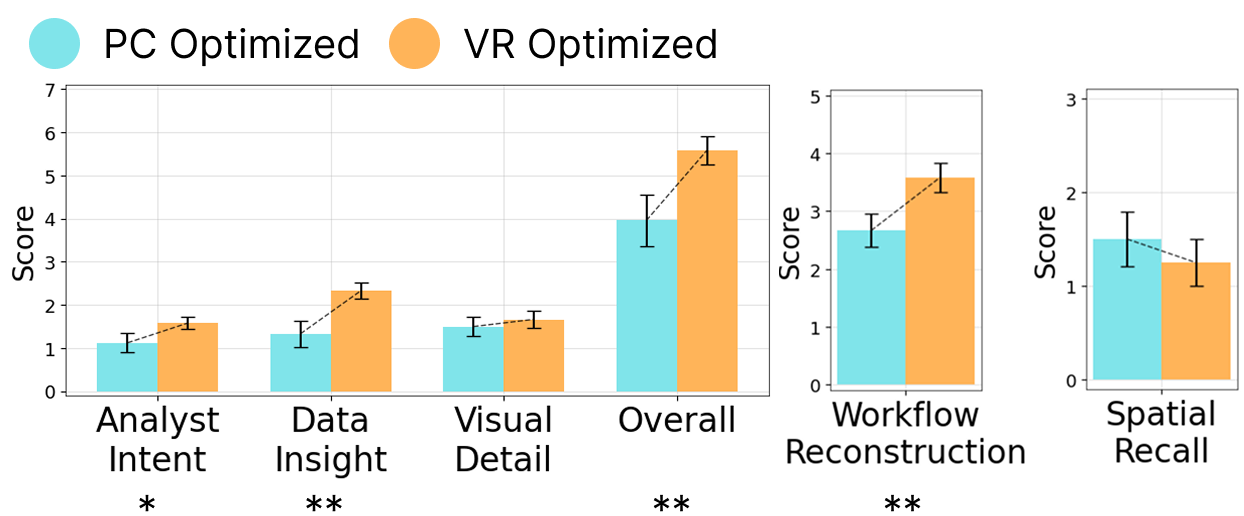}
    \caption{ Performance Comparison in Optimized PC and VR Condition: Average Comprehension Sub-scores, Reconstruction Score and Spatial Recall }
    \label{fig:fig3}

    \vspace{2.1\baselineskip} 

    \includegraphics[width=1\linewidth]{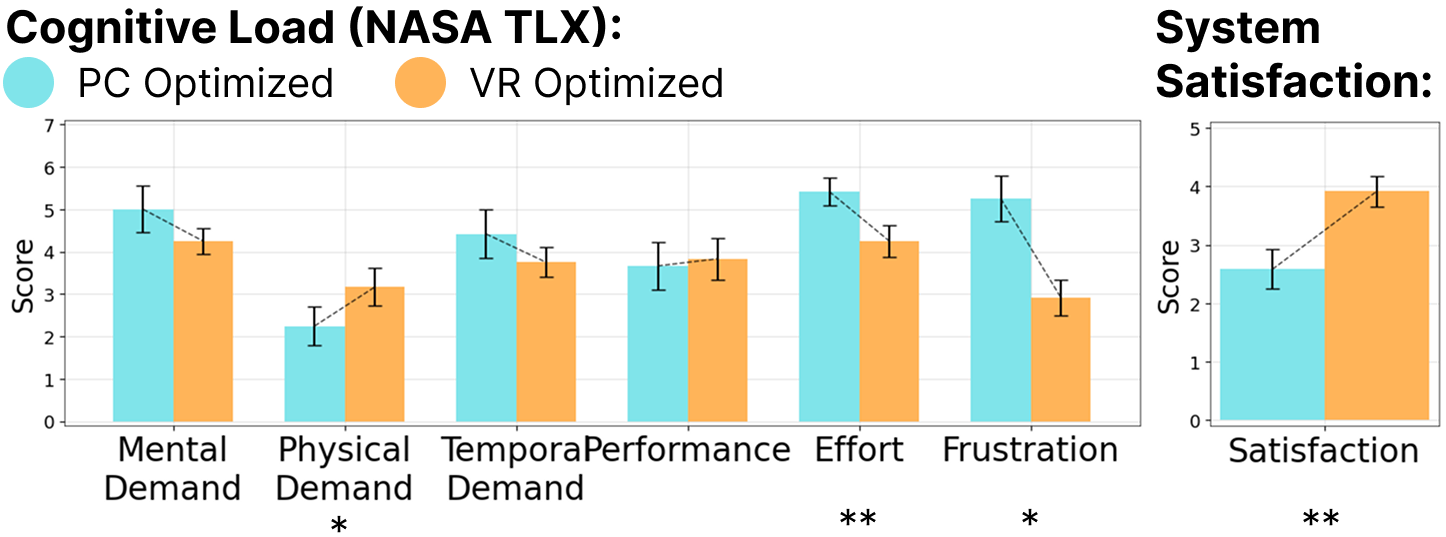}
    \caption{Subjective Ratings Comparison for Optimized PC and VR Condition: Workload and Satisfaction }
    \label{fig:fig4}
\end{figure}

\textbf{Task Performance:} As shown in \autoref{fig:fig3} Our analysis revealed significant differences in task performance between the PC and VR conditions, as measured by comprehension, workflow reconstruction, and spatial recall scores. In support of $H(Perf)$, the VR condition led to significantly better task comprehension. A Wilcoxon signed-rank test showed that participants achieved a higher mean comprehension score in VR (\textit{M} = 5.58, \textit{SD} = 1.12) compared to PC (\textit{M} = 3.96, \textit{SD} = 2.08), with the difference being statistically significant (\textit{W} = 6.5, \textit{p} = .018). Further supporting $H(Perf)$, the VR condition also resulted in superior workflow reconstruction accuracy. Participants' workflow scores were significantly higher in VR (\textit{M} = 3.58, \textit{SD} = .90) than in PC (\textit{M} = 2.67, \textit{SD} = .98), as indicated by a Wilcoxon signed-rank test (\textit{W} = 4.00, \textit{p} = 0.047).

However, we did not find a significant difference in spatial recall scores between the VR (\textit{M} = 1.25, \textit{SD} = .87) and PC (\textit{M} = 1.50, \textit{SD} = 1.00) conditions (\textit{W} = 9.50, \textit{p} = .44), partially rejecting $H(Perf)$. A follow-up analysis did reveal that within the VR condition, participants with greater data visualization experience tended to have higher spatial recall scores, showing a moderate positive correlation (\textit{r} =.56), using Spearman's rank correlation.

\textbf{Subjective Ratings:} As shown in \autoref{fig:fig4} We assessed cognitive and physical load using the NASA-TLX survey. Our findings partially support \textbf{$H(Load)$:} There were no statistically significant differences in the mental, temporal, or performance demand subscales between the two conditions. However, regarding physical demand, VR (\textit{M} = 3.17, \textit{SD} = 1.53) yielded significantly higher scores than PC (\textit{M} = 2.25, \textit{SD} = 1.60) using Wilcoxon signed-rank test (\textit{W} = 8.0, \textit{p} = .07). 

The PC condition was perceived as significantly more frustrating than the VR condition (\textit{W} = 11.5, \textit{p} = .05). In line with $H(Pref)$, a Wilcoxon signed-rank test on the satisfaction scores revealed that participants found the VR replay to be more satisfying than the PC replay (\textit{W} = 11.5 , \textit{p} = .03). Furthermore, when asked to rank their preference, a vast majority of participants (10 out of 12) chose the VR condition as their preferred format.

\begin{figure}
    \centering
    \includegraphics[width=1\linewidth]{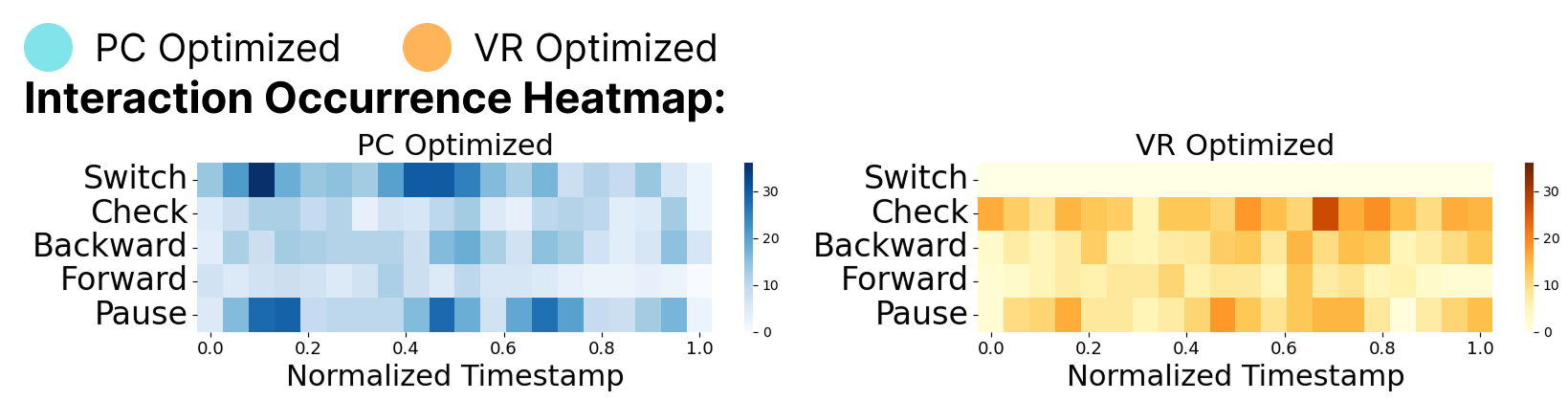}
    \caption{Heatmaps of interaction counts across the normalized task timeline for tested Conditions. Darker colors represent a higher frequency of user interactions within that time window.}
    \label{fig:heat}
\end{figure}

\textbf{Behavioral Data:}
\added{We examined the difference of behavioral metrics, including task completion time, pause duration, pause counts and perspective switch counts.} While no significant difference was detected, \autoref{fig:heat} provided insights into the user interaction frequency. The darker areas signifies more frequent interactions, suggesting critical periods in the replay. Besides, on PC, higher proportion of 1PP duration was positively correlated with better comprehension score (Spearman's \textit{r} = 0.61, \textit{p} = .04).

\added{\textbf{Qualitative Feedback:}
A thematic analysis was adopted to synthesize feedback into three main topics: \textit{1. perspective}, \textit{2. new features} and \textit{3. sensemaking strategy}.} 

\added{\textit{\textbf{Agency and visual clarity drove a preference for VR}}, which was consistent with $H(Pref)$. Eight out of 12 participants cited freedom of perspective as the primary reason, which not only allowed them to look from desired angles (P3, P5, P11) but also offered clear visual information (P3, P4, P5, P8, P9). In contrast, the PC condition was criticized for its limited perspective; participants noted that while the avatar no longer hindered the view, visual overlaps still occurred when multiple visualizations existed (P6) and the 3PP view made it difficult to investigate details in 3D scatterplots (P7). } 

\added{\textit{\textbf{New interactive features functioned as core cognitive tools for information processing.}} The ability to pause and rewind was acknowledged by ten participants as an essential feature. P4 and P7 also mentioned using it as a core part of their strategy to process information. The proximity-based avatar fade-out was a subtle but effective improvement that successfully prevented the avatar from occluding critical data (P5, P9). Regarding navigation, the perspective switch was actively used. Most particpipants found it smooth to use, but some described it as not ideal (P1, P6).}

\added{\textit{\textbf{Participants adopted active sensemaking strategies to manage analytical complexity.}} Two distinct strategies emerged. The most prevalent approach used by six participants was to go over the entire replay first. Alternatively, some participants (P7, P11) adopted a layered strategy, watching the entire replay once for a general overview before using timeline controls to go back to specific moments for details. These behaviors demonstrate that the replay system supports personalized, active reconstruction of the analytical process rather than passive observation.}

\subsection{Discussion}

The results of Study 2, which compared the optimized replay formats for PC and VR, provide clear evidence supporting the immersive, embodied approach for re-experiencing complex analytical workflows. The discussion below interprets these findings, focusing on the fundamental differences between the two media and the role of the newly implemented interactive features.

\textbf{Embodied Agency vs. Guided View:} Phase 1 illustrated that guided view was an effective practice on PC, while embodied agency was critical for VR. The meaningful comparison in Phase 2, therefore, was to determine which of these optimized paradigms would better support complex IA replays. Our results show a clear advantage for the VR condition in task comprehension (\textit{p} = 0.018) and workflow reconstruction (\textit{p} = 0.041), suggesting that for understanding complex analytical processes, the benefits of embodied agency are more critical than those of a guided view.

The ability to navigate and control one's perspective in VR allows for the construction of a robust mental model of the complete data history. This aligns with findings from Chrastil \& Warren\cite{chrastil2015active}, which suggest that self-directed navigation is superior for acquiring cognitive map knowledge. Moreover, the task of analyzing a 3D data visualization is intrinsically spatial. The depth dimension in VR provides a more direct representation of the data, reducing the cognitive effort needed to interpret 3D structures on a 2D screen.

\textbf{Effectiveness of pause, rewind and perspective switch:} 
\added{Most participants believed pause and rewind was the most preferred feature. Participants explicitly described using the pause function to ``process information'' and ``think'' (P4), suggesting a direct mitigation of the high temporal demand. }By giving users control over the pace of information flow, these features allowed them to manage their own cognitive load. On the other hand, for perspective switch, while most users do it multiple times and think of it as positive, we do not find correlation between perspective switch counts and task performance. 
Moreover, a Spearman's rank correlation identified the proportion of time spent in 1PP to be positively correlated to comprehension score (\textit{r} = 0.608, \textit{p} = 0.0359), which further implies that 3PP view in PC tend to miss details in visualised data, due to less guided focus and longer distance.

\textbf{Why spatial recall indicated no signicant difference between PC and VR:}
% We present a possible explanation for this lack of salience. 
\added{This result partially rejects hypothesis \textbf{$H(Perf)$}, suggesting that an immersive environment does not automatically confer better spatial memory for all users.} Our demographic analysis provides a critical insight here, as prior experience with data visualization showed a moderate positive correlation with spatial recall performance (\textit{r} = 0.56) only in the VR condition. This implies that the spatial advantages of VR are not automatic; instead, they are most effectively leveraged by users who already possess well-developed spatial reasoning skills. This finding resonates with literature suggesting that individual differences can moderate performance in virtual environments \cite{iriye2021memories}.

%% file: sections/06-study3.tex
\section{Design Guidelines and Implications for IA Replay}
With the insights from the two studies, we try to convert the practical lessons from our work into four concrete, implementable guidelines for designers of IA replay systems. Notably, while we concluded VR as the better performing platform, there are many scenarios where only traditional desktop settings are available, especially for asynchronous, remote collaboration. Therefore, we try to target both PC and VR in the design guidelines we present.

\textbf{Match navigation to the medium: give VR agency, keep PC guided.}
A primary takeaway from our work is that the optimal user perspective and navigation control is fundamentally tied to the medium. In VR, the replay system should let viewers physically reposition themselves or use natural locomotion to inspect visual elements from different angles and distances. \added{It is an overarching guideline to preserve the users' agency for navigation, so as to immerse users in the analytical environment for better performance and ensure usability. 
These observations align with previous work finding physical navigation outperform virtual navigation~\cite{ball2007move,in2024evaluating}.}
On the contrary, for desktop platforms, where users view a 3D visualisation scene through a 2D window, the priority should be a guided and focused perspective. 
Navigation effort should be minimized, with precurated camera paths, as manual navigation increased cognitive load without measurable performance gains. 
\added{Such navigation burden were also reported in previous studies~\cite{enriquez2024evaluating,tong2023}.}
The perspective change feature needs to be implemented with caution, as it may lead to loss of focus. It should be considered for coarse-grained, procedural tasks instead of fine-grained, detail-rich analysis.

\textbf{Reduce visual occlusion and increase clarity.}
Common obstacles across both studies were the avatar or other visuals blocking important information and users struggling to keep pace with replayed timelines. 
\added{Implementations such as proximity-based avatar fading~\cite{elmqvist2008taxonomy}, pause/rewind, and timeline slider proved to be useful.}
These features are easily overlooked, but they could be the key point to user experience and task  comprehension, as shown in the comparison of 1PP and 3PP Passive conditions.
\added{Research indicates that while 1PP enhances immersion and precision, 3PP significantly aids spatial awareness and context gathering during task review~\cite{salamin2010quantifying,ueyama2022effects}.
Consequently, we recommend treating these not as optional add-ons, but as baseline requirements for IA replay systems and VR usability guidelines.}
To further reduce visual overlap, it is beneficial to keep track of the visualization being interacted with or investigated, based on proximity or interaction, \added{a concept aligned with Focus+Context visualization techniques~\cite{kalkofen2007interactive,yang2022pattern}.}
The tracked visualization could then be highlighted or zoomed in to provide detail. Other visual elements could also reduce opacity to further emphasize the targeted visualization, ensuring that contextual information is preserved without creating visual clutter.

\textbf{Support replay as an active sensemaking tool instead of a passive video.}
Our studies showed that participants rarely treated replay as a passive video. 
\added{Instead, they actively managed the pace and sequence of playback to support their own reasoning, a behavior consistent with the ``active viewing'' strategies observed in interactive video learning~\cite{merkt2011learning}.}
In Study 2, the pause and rewind feature was described as ``most important'' and became a core strategy. Participants frequently paused after a key analyst action to consolidate their understanding or used rewind to revisit details they had missed. Some also reported deliberately watching the entire replay once to gain an overview, then returning to specific points to answer detailed questions. 
\added{This specific workflow mirrors Shneiderman's Visual Information-Seeking Mantra: Overview first, zoom and filter, then details-on-demand~\cite{shneiderman2003eyes}, confirming that replay is already functioning as an analytic instrument rather than a static recording~\cite{ragan2015characterizing}.}
Building on this, future systems should strengthen replay’s role in collaborative sensemaking by tying replay more directly with analytic actions, moving towards analytic provenance~\cite{ragan2015characterizing}. 
For example, segments could be linked to operations such as construct, filter, or compare~\cite{gotz2009characterizing}, allowing collaborators to navigate by action sequence rather than just timestamps.
Similarly, key frame function could make it easier for analysts to capture and share the key steps they pause to reflect on. In this way, replay evolves from a passive narrative into an interactive workspace for reconstructing and communicating analytic provenance.

\section{Limitations and Future Works}
\textbf{Annotation and Note-Taking:} Our investigation focused on the core parameters of the replay experience through the two-phase controlled study. For our purpose, we did not test all potentially beneficial features, most notably the integration of explicit insight communication. While our replay system captures the procedural workflow, the original analyst's specific ``aha'' moments and rationale remain implicit. Future immersive analytics replay systems could benefit from techniques such as audio/video annotation~\cite{irlitti2013tangible}, text~\cite{dominic2020exploring, lee2019annotation, ulusoy2018beyond} and highlights~\cite{mahmood2019improving, reski2020oh, ulusoy2018beyond}.  Investigating the most effective methods for embedding these insights directly into the replay could significantly enhance a collaborator's understanding of the analyst's intentions. 

\added{\textbf{Navigational and Temporal Interactions:} Another limitation lies in the navigational guidance and temporal control mechanisms, especially for long-duration replay sessions. While active navigation in VR provides agency, users may struggle to identify where to look, increasing cognitive load and the risk of missing critical visual details. Future systems could address this by incorporating visual attention cues such as gaze rays~\cite{chen2023gazeraycursor} or head rays~\cite{piumsomboon2019effects}, in order to indicate avatar gaze and highlight visualizations currently being manipulated, and thereby guide viewers without sacrificing their sense of agency. To mitigate the physical demand and fatigue noted by participants in extended sessions, future systems can incorporate teleportation~\cite{buttussi2019locomotion, prithul2021teleportation} or ``follow analyst'' re-centering features~\cite{nguyen2020human} to provide proactive guidance while maintaining comfort.}

\added{Furthermore, to allow more effective  timeline control, future work should explore novel temporal interfaces, such as allowing users to place their own keyframe marks~\cite{zhao2017supporting} for efficient navigation to critical moments. Additionally, more precise slider interaction techniques~\cite{perelman20253d,zhang2025forcepinch} could be integrated to provide finer temporal control when analyzing dense sequences of data interactions.}

\added{
\textbf{Semantic Filtering and Clipping:} Inherently, raw replays of analytical processes are noisy, often containing idle time and locomotion that are irrelevant to the data analysis. Removing these segments would make the replay more focused and time-efficient, as watching lengthy replays imposes a significant cognitive burden on the viewer. Future systems could address this by implementing automated filtering and clipping based on key periods identified in provenance logs and user behavior~\cite{zhang2025summact}. Furthermore, the integration of LLM-assisted analytics frameworks~\cite{kim2025explainable} could help transform complex interaction sequences into a structured, concise narrative. Vision Language Models help remove noises through comprehending and clipping lengthy session recordings~\cite{shu2025video}. By automatically pruning idle periods and highlighting high-value analytical events, these techniques can lower the burden of session summarization, allowing incoming analysts to rapidly grasp a predecessor's rationale during a handover.}

%% file: sections/07-conclusion.tex
\section{Conclusions}

In this paper, we presented an empirical study evaluating session replay techniques for asynchronous \added{task handover} in immersive analytics. We evaluated the impact of three key design dimensions: media, perspective, and navigation control, through a two-phase within-subject comparison. Our results reveal that the optimal design for such systems is highly platform-dependent. For VR, a third-person perspective with free, active navigation significantly improved task comprehension and workflow reconstruction. Conversely, on PC, A passive, guided experience was more preferable. A direct comparison of the two optimized formats confirmed that the immersive VR replay led to significantly better task completion time and comprehension accuracy. Our findings can inform the design of future asynchronous collaborative systems in data-centric immersive environment.